\documentclass[12pt,preprint]{aastex}

\shortauthors{T. Hosokawa}
\shorttitle{Constraining the Lifetime of QSOs}

\begin{document}

\title{Constraining the Lifetime of QSOs with Present-day Mass
Function of Supermassive Black Holes} 

\author{Takashi HOSOKAWA}
\affil{Yukawa Institute for Theoretical Physics, Kyoto University,
Sakyo-ku, Kyoto 606-8502, JAPAN}
\email{hosokawa@yukawa.kyoto-u.ac.jp}

\begin{abstract}

Using the theoretical models of the QSO formation,
we can reproduce optical QSO luminosity functions (LFs)
at high redshifts ($z \geq 2.2$).  
Two different models can reproduce LFs successfully, though the lifetime
of QSOs, $t_Q$, and the relation between the black
hole mass and the host halo mass are different each other;
$t_Q \sim 10^6$yr, in one model, $t_Q \geq 10^7$yr, in other models.  
Here, we propose a method to break this degeneracy.   

We calculate the mass function of supermassive black holes (SMBHs)
at $z=2.5$, and compare the result with the current mass function obtained
by Salucci et al.(1999).
In the shorter lifetime model, the mass function at $z=2.5$ exceeds that of
$z=0.0$ by one order of magnitude, then it should be ruled out.
We conclude that the lifetime is at least $t_Q \geq 10^7 {\rm yr}$. 
  
Next, we examine the difference 
of the formation epoch of SMBHs existing at $z=3.0$ for each model under the
model assumptions. We simply discuss the difference of formation epoch as 
another possible model-discriminator.
   
\end{abstract}


\section{Introduction}

Today, we observationally know that the population of QSOs evolves on
the cosmological timescale. At $z \sim 2-3$, 
the comoving QSO number density was at maximum, which is 
$\sim 1 {\rm QSO}/(100 {\rm Mpc})^3$. At $z<2$, 
comoving number density of QSOs drops down quickly toward $z \sim 0$,
when the number density of bright QSOs are only 1/100 compared with 
that of $z \sim 2.5$.
This property is observed in optical (Boyle \& Terlevich 1998), 
soft X-ray(Miyaji et al.2000), and radio(Shaver et al. 1998).
At $z > 3$, on the other hand, the change of the QSO number density 
is different in different wavebands.
In optical and radio bands, QSOs decrease slowly beyond $z \sim 3$, 
there were 1/10 bright QSOs 
at $z \sim 4.5$ compared with the number at $z \sim 2.5$
(e.g. Fan et al. 2000, Shaver et al.1999 ). 
This slow down pace agrees between optical and radio observation.
In soft X-ray, on the other hand, the number density is approximately
kept  constant beyond $z \sim 3$ according to the observation with 
ROSAT (Miyaji et al. 2000).
 
In our universe, the structures, such as galaxies, is thought to
be formed from initial density fluctuation through the gravitational
instability. In this scenario, the most of the matter must be non-baryonic
, what is say, dark matter (DM). Today, various properties of the 
large scale structure
can be explained in terms of the cold dark matter(CDM) model.
QSOs are active nuclei of galaxies, therefore, it is natural 
that the evolution of QSO population reflects the structure formation
history under the CDM cosmology.
Based on such motivation, many authors (Haehnelt et al. 1994,
Haiman \& Loeb 1998, Haehnelt et al.1998, Hosokawa et al.2001
,hereafter Paper I) model successfully QSO 
population at high redshifts by using the formalism based on 
Press-Schechter theory (Press \& Schechter 1974).
The reason why these studies are limited to high 
redshifts is that the Press-Schechter theory cannot directly treat the 
substructures of halos.
Treating QSOs, the relevant mass scale of dark halos should be galactic halo
scale, $\sim 10^{11}-10^{13} M_\odot$, and at low redshifts, a lot of these
halos exist as the substructures of more massive halos whose scale
is more than galactic cluster, $\sim 10^{14} M_\odot$.
In these studies, modeling of QSOs may be oversimplified, but thanks to
its simplicity itself, it is easy to use this QSO population model to
investigate the role of QSOs at high redshifts; for example,  
the reionization (Haiman \& Loeb 1998),
X-ray background (Haiman \& Loeb 1999), and so on.
Furthermore, QSOs often become the source objects of the gravitational
lens, because of its cosmological distance. 
When we consider lens statistics with QSOs, these models are useful
(Bartelmann 2000).

To understand the reason of the rapid decline of QSO population
at $z < 2$, the full semi-analytic
approaches of the galaxy formation which incorporate the QSO
activity and the feeding supermassive black holes(SMBHs) were adopted
(e.g. Kauffmann \& Haehnelt 2000, Haehnelt \& Kauffmann 2000,
 Monaco et al. 2000).  
Kauffmann \& Haehnelt(2000) and Haehnelt \& Kauffmann (2000) 
successfully explain the optical luminosity
functions(LFs) at low redshifts and the relation of nearby galaxies 
between the mass of dormant SMBHs and its bulge properties(e.g. Magorrian
et al. 1998, Ferraresse \& Merritt 2000). 
However, in both of the full semianalytic models and simple toy models at high
redshifts, the physical meanings of modeling the QSOs are poor.
We need to check the justification of QSO models in terms of the 
observations to understand its meanings.
 
In this paper, we employ the simple model at high redshifts.
The formation process of dark halos is straightforward once
the cosmology and the DM model are given. 
In these model, the model parameter is ``the QSO lifetime'', $t_Q$,
where QSO shines at Eddington luminosity, and the relation between
the mass of SMBHs, $M_{\rm BH}$, and that of host halos, $M_{\rm halo}$.
Haehnelt et al.(1998) tests two types of models, linear relation
between $M_{\rm BH}$ and $M_{\rm halo}$, power law relation suggested
by Silk \& Rees(1998). Actually, both of these models can reproduce
QSO LFs at $z \sim 3$. 
At this time, in the former case, QSO lifetime is shorter, $t_Q \sim
10^6$yr, in the latter case, QSO lifetime is longer, $t_Q \sim 10^7$yr.
The cause of this degeneracy is the balance
between QSO lifetime and the fraction $M_{\rm BH}/M_{\rm halo}$.
Haiman \& Hui(2001) and Martini \& Weinberg(2001) present the
diagnosis to break this degeneracy using the QSO clustering.
In this paper, we suggest an alternative way to discriminate models, 
and will show that the shorter lifetime model should be ruled out.

Below, we describe the content of models and represents that these
models successfully reproduce optical LFs at high redshifts($z>2.2$).
Next, we show that the shorter lifetime model should be ruled out
based on the calculations of the mass function of SMBHs.
Finally, we investigate the distribution of the formation epoch of SMBHs
existing at $z=3$, and discuss a possibility to resolve the models.

\section{Reproduced QSO LFs and Model Parameters}

First, we reproduce the optical QSO luminosity functions (LFs) at
high redshifts, $z \sim 2.6,3.0$, and $4.4$.
In Paper I, we fit both of the optical and the X-ray 
LFs only at $z \sim 3$. Here, we aim at reproducing the 
optical LFs at a wider range of high redshifts.

Our model is very simple and based on the works by 
Haehnelt et al. (1998) and Haiman \& Loeb (1998).
Since the adopted models in this paper are almost the same as 
those of Paper I, we only briefly explain them here. 
Since the formation process of QSOs is unknown, the basic
scenario is needed.
We assume that the each dark halo necessarily possesses one SMBH,
and that feeding SMBHs occurs immediately after the host halo has formed.
That is, QSOs begins to shine as soon as its host halo have formed. 

To reproduce QSO LFs, we assume the following relations
between the mass of the SMBH and host halos,
\begin{equation}
({\rm A}) \ \ \ M_{{\rm BH}} = C v_{{\rm halo}}^5 
                =C'M_{{\rm halo}}^{5/3} \times (1 + z_f)^{5/2}  
\label{modela}
\end{equation}
\begin{equation}
({\rm B}) \ \ \ M_{{\rm BH}} = \epsilon M_{{\rm halo}} 
\end{equation}
Here, $z_f$ is the redshift when the host halo forms. 
The $z_f$ dependence in model A has been introduced through the relation
$v_{\rm halo}^5 \propto (M_{\rm halo}/r)^{5/2}
\propto M_{{\rm halo}}^{5/3} ~ \rho_{{\rm halo}}^{5/6} 
\propto M_{{\rm halo}}^{5/2} (1+z_f)^{5/2}$, and its
physical basis is given by Silk and Rees(1998). Namely, equation 
(\ref{modela}) gives the upper limit of black hole mass to bound the 
gas to feed SMBHs against feedback from the QSO.
If one can check equation (\ref{modela}), 
the growth of SMBHs must have been controlled by intense radiation of QSOs.   
In model B, we simply extend the linear relation between
the BH mass and the bulge mass found by the observation of
nearby galaxies (e.g. Maggorian et al. 1998, Meritt \& Ferrarese 2000). 
But here, the relation of model B does not contain 
the bulge mass but the halo mass.  
In comparison, we consider the third model,
\begin{equation}
({\rm C}) \ \ \ M_{{\rm BH}} = C''M_{{\rm halo}}^{5/3}. 
\end{equation}
This model is similar to model A but does not contain the dependence on 
the formation epoch $z_f$. 
Our motivation for adopting this relation 
is to examine the importance of the $z_f$ dependence in model A, 
Although the difference will be shown not to be large
in reproducing LFs(see below).

Since we assume that each dark halo possesses only one SMBH, 
the formation rate of SMBHs can be derived from that of 
dark halos. The methods of calculating the formation rate  
of dark halos based on the excursion set approach 
have been proposed by many authors (e.g. Lacey and Cole 1993,
Kitayama and Suto 1996).
We adopt the one proposed by Kitayama \& Suto, 
in which the formation rate of black holes is given by 
\begin{equation}
\label{Form}
\frac{d^2N_{\rm form}}{dM_{\rm BH}dz}(M_{\rm BH},z_f,z) = 
\frac{1}{\epsilon}
 \frac{d}{dz} \frac{dN_{\rm form}}{dM_{\rm halo}}(M_{\rm halo},z_f)
  \times P_{\rm surv}(M_{\rm halo}|z_f,z), 
\end{equation}  
where, $d(dN_{\rm form}/dM_{\rm halo})/dz$
is the formation rate of dark halos and $P_{\rm surv}(M_{\rm halo}|z_f,z)$ 
is the {\it survival probability}, the probability 
that the dark halo of $M_{\rm halo}$ which was formed 
at $z_f$ remains at $z$ without merging into objects of higher masses.  
Kitayama and Suto calculated the genuine formation rate 
of the halo with mass $M$ by
the merging rate of halos of $< M/2$ to create a halo of mass $M$.

We next assume that
the time evolution of the QSO luminosity follows
\begin{equation}
\label{L}
 L(t) = L_{\rm Edd} \exp \left( - \frac{t}{t_{\rm Q}} \right)
       \equiv M_{\rm BH}~g(t),
\label{lumi}
\end{equation}
where we set $t=0$ when a halo collapses. 
Such a simple prescription for a single
QSO light curve is known to well reproduce
the observed LF and thus has been used frequently
(e.g. Haiman \& Loeb 1998, Haehnelt et al. 1998, 
Kauffman \& Haehnelt 2000, Paper I).

Finally, we calculate the QSO LF by the summation of the luminosities of
all the QSOs whose luminosity is $L$ at redshift $z$, that is
\begin{eqnarray}
\label{LF}
 \Phi(L,z) \!\!\!&=&\!\!\! 
               \int^\infty_0 \int^z dM_{\rm BH} dz_f \
               \frac{d^2N_{\rm form}}{dM_{\rm BH}dz} 
               \ (M_{\rm BH},z,z_f)    \nonumber\\
        \!\!\!&\times& \!\!\!
	        \delta \left[M_{\rm BH}-\frac{L}{g(t_{{z_f},z})}\right],
\end{eqnarray}
where $\delta$ is the delta-function,
$g(t)$ is defined in equation (\ref{L}), and
$t_{z_f,z}$ is the time between the epochs of redshifts $z_f$ and $z$.
To compare the theoretical model with the observation, we need to
convert $\Phi(L,z)$ to the LF in the observational band
$\Phi(L_{\rm band},z)$. As in Paper I, we use the 
QSO model spectrum calculated based on the disk-corona model by 
Kawaguchi et al.(2001) (See Fig.1 in Paper I).

The best-fit QSO LFs at high redshifts are shown in Fig.1,2,and 3, 
together with the observed LFs for model A, B, and C respectively.
Here, we adopt $\Lambda$CDM cosmology model and 
the adopted cosmological parameters are
$(\Omega_0, \Omega_\Lambda, \Omega_{\rm b}, h, \sigma_{8h^{-1}}, n)
= (0.35, 0.65, 0.04, 0.65, 0.87, 0.96)$.
The observational data at $z = 2.6$ and $3.0$ are given by
Pei(1995). We add the data by Schmidt et al.(1995), Kennefick et al.(1995), 
Kennefick et al.(1996) for the LFs, at $z=4.4$. 
The data by Pei(1995) are given under 
$(h,q_0, \alpha)=(0.5,0.5,0.5),(0.5,0.1,1.0)$, 
and the data at $z=4.4$ by Kennefick et al(1996) and others are given under
$(h,q_0, \alpha)=(0.5,0.5,0.5)$, 
where $\alpha$ is the spectral index defined by 
$L_\nu \propto \nu^{- \alpha}$. The dependence of $\alpha$ derives 
from K-correlation. Now, the situation is complex since our SED depends on
$M_{\rm BH}$ and $\dot{M}$ (see Paper I); that is, $\alpha$ is not constant.
In Paper I, we adopt $\alpha=1.0$ data of Pei(1995)
, where there is no K-correration since $\nu L_\nu = {\rm const.}$ and
we calculate the LFs in the redshifted waveband.
Here, however, we cannot simply calculate LFs at $z=4.4$ 
in redshifted waveband
since the Lyman limit(912 \AA) redshifts to longer wavelength than
B-band(4400 \AA). Furthermore, it is unreasonable to use the data of
$\alpha=1.0$ and $\alpha=0.5$ together.
Thus, we adopt the data of $(h,q_0,\alpha)=(0.5,0.5,0.5)$ at all redshifts.
These data are properly modified, following the rule 
$\Phi(L,z) \propto dV^{-1} d_L^{-2}$, $L \propto d_L^2$,
where, $dV$ is the volume element, and $d_L$ is the luminosity distance.
 
As shown in Paper I(see also Haehnelt et al. 1998, or Haiman \& Hui 2001),
best-fit parameters which can fit observational QSO LFs  
differ in different models. Now, the values of parameters are
\begin{itemize}
\item Model A \ ; \ $t_Q = 6.0 \times 10^7$ yr¡¢
$M_{\rm BH} = 1.58 \times 10^9 \times (v_{\rm halo}/500 {\rm km/s})^5$
\item Model B \ ; \ $t_Q = 1.0 \times 10^6$ yr¡¢
$M_{\rm BH}/ M_{\rm halo} = 1.26 \times 10^{-3}$
\item Model C \ ; \ $t_Q = 1.0 \times 10^7$ yr¡¢
$M_{\rm BH} = 7.13 \times 10^{-13} \times M_{\rm halo}^{5/3}$
\end{itemize} 
The best fit values of $t_Q$ are very different among these models
by more than one order of magnitude. Despite this difference,
all the models can well reproduce the optical LFs.
This is because of the different ratios being assigned, 
$\epsilon = M_{\rm BH}/M_{\rm halo}$. 
This is clear by Fig.4.
Fig.4 represents the minimum SMBH mass and halo mass
in the calculation of the QSO LFs at $z=3.0$.
Our formalism(equation (\ref{LF})) to calculate LFs is
integrating the more massive SMBHs and halos, 
from the redshift $z$ to infinity. 
But the dominant population of SMBHs and halos are those 
formed just before $z$ since more massive population formed earlier
is minority in CDM cosmology.
This shows that the more massive halos and
SMBHs contribute to LFs on the more luminous side. 
And this figure shows that the dominant halo mass scale contributing to
LFs at $L_B \sim 10^{12} - 10^{14} L_{B,\odot}$, where we compare
with the observational data, is $M_{\rm halo} \sim 10^{12.5} -
10^{13.5} M_\odot$. This is just the galactic halo mass scale. 

One can see the effects of different $\epsilon \sim M_{\rm BH}/M_{\rm halo}$ 
in each model. Namely, when $t_Q$ is shorter,
the value of $M_{\rm BH}/M_{\rm halo}$ should be larger to fit the
observed LFs.
Since in our model, the initial luminosity of QSOs is assumed to be 
the Eddington luminosity, large $M_{\rm BH}/M_{\rm halo}$ 
means that each halo possesses more luminous QSO than otherwise. 
After all, QSOs shine longer in model A than in model B,
but each QSO is less luminous since Eddington luminosity
is proportional to $M_{\rm BH}$.

Same as paper I, we can calculate QSO LFs in soft X-ray band with the
same parameters. However, we note that the calculated LFs at $z \sim 3$
exceeds the observational data (Miyaji et al. 2001) by about
one order of magnitude.
This is because we firstly reproduce the optical LFs in the wide range 
of redshifts in this paper. However, many studies (e.g. X-ray background;
Gilli et al. 2001 or BAL QSOs; Brandt et al.2000)   
suggest the existence of the X-ray absorbed QSOs. 
Or at high redshifts, a lot of QSOs shining at nearly Eddington luminosity
may have steeper X-ray spectrum such as narrow-line Seyfert galaxies. 

As mentioned above,
at lower redshifts than $z \sim 2.5$, this simple model can
no longer fit the observed LFs. 
Haiman and Menou(2001) noted that the redshift distributions of 
merger between dark halos of $\sim 10^{12} M_\odot$ have a
peak around $z \sim 2.5$. 
That is to say, the galactic clusters begin to form at $z \sim 2.5$.
Generally, however, we cannot directly treat subhalos in the dark halos
in the framework of Press-Schechter formalism(sub-halo problem),
and the assumption that 
one dark halo possesses only one SMBH becomes unreasonable at lower redshifts.
The full semianalytic treatment(Kauffmann \& Haehnelt 2000) or 
the interaction in the galactic cluster(Cavaliere \& Vittorini 2000) 
can explain this steep declination.
Nevertheless, our model seems to be a good approximation at $z \geq 2.5$,
we consider SMBHs at higher redshifts using this model in the following.

\section{Observational Possibility to Break the Degeneracy}

As mentioned above, we cannot determine the most suitable model
from fitting to the observed QSO LFs.
Haiman and Hui(2001) and Martini and Weinberg(2001) argue that 
from the observations of QSO clustering we will be able to get 
information regarding the QSO lifetimes.
In this section, we propose an alternative way to break this degeneracy.

\subsection{Mass Function of SMBHs}

Haiman \& Hui (2001) briefly evaluate the maximum $M_{\rm BH}/M_{\rm halo}$
based on $M_{\rm BH}/M_{\rm bulge} \sim 0.006$ given by Maggorian et al.(1998)
and conclude that the minimum QSO lifetime is $t_Q \geq 3 \times 10^6 
{\rm yr}$ in order to fit the observed QSO LFs. 
Here, we calculate the mass function of dormant SMBHs at $z=2.5$, and
compare with the present-day mass function given by Salucci et al.(1999).
Salucci et al. used smaller ratio $M_{\rm BH}/M_{\rm bulge} \sim 0.002$,
then the constraint should be more stronger.
Based on these models in section 2, we calculate 
the mass function of dormant SMBHs at $z=2.5$; that is
\begin{equation}
\frac{d \Phi}{d M_{{\rm BH}}} (M_{\rm BH},z=2.5) 
= \int^z_{z_{\rm max}} \frac{d^2 N_{\rm form}}{dz dM_{\rm BH}}
(M_{\rm BH}, z_f,z)  ~dz_f .
\label{eq:bhmf}
\end{equation} 
Here, $z$ is the redshift where we consider the mass function of SMBHs,
and $z_{\rm max}$ is the maximum redshift for the integration,
which we set $z_{\rm max} = 15.0$.
The BH formation rate, $d^2 N_{\rm form}/dM_{\rm BH} dz$
is given by equation (4). 
Among these models, $\epsilon$ depends on $z_f$ only in model
A, otherwise $\epsilon$ is constant.

Fig.5 represents calculated BH mass function 
at $z=2.5$ in the mass range of 
$10^5 M_\odot \leq M_{\rm BH} \leq 10^{11} M_\odot$.
The corresponding masses of the host halo for the same SMBH 
are different in each model. 
This reflects the different $M_{\rm BH}/M_{\rm halo}$ ratios 
as has mentioned above.
We should set the lower limit of the mass of the halos which can 
possess SMBHs. Menou et al.(2001) use the redshift-dependent relation 
given by Navarro, Frenk, \& White (1997),
\begin{equation}  
M_{\rm halo} \geq 9 \times 10^7 M_\odot 
\left( \frac{\Omega_0}{0.3} \right)^{-1/2}
\left( \frac{1+z_f}{10} \right)^{-3/2}
\left( \frac{h}{0.7} \right)^{-1} .
\label{eq:lowm}
\end{equation}
This is the condition for  $T_{\rm vir} \geq 10^4$K, since 
otherwise the radiative cooling with atoms is not efficient, thus, the gas
cannot accrete to the center.  
That is, this represents the lower limit of $M_{\rm BH}$ for
baryons to cool efficiently and accrete to SMBHs.
At $z_f \geq 2.5$, equation (\ref{eq:lowm}) gives 
$M_{\rm halo,min} \sim 10^8 M_{\odot}$.
  
As mentioned above, the BH mass corresponding to one host halo mass
is very different for different models. 
Mass of the host halo for given BH mass is lowest in model B. 
Since the number density of halos are a decreasing function of halo mass,
mass function of model B is larger than others by 1 or 2 orders
of magnitude.

Using these results, we can constrain the models.
Since it is difficult to know the mass function of SMBHs at $z \sim 2.5$;
we use the estimate of mass function at the present-day, provided
by Salucci et al.(1999). They analyze nearby galaxies
in two methods. First, they simply employ the $M_{\rm BH}-L_{\rm bulge}$
relation(e.g. Merritt \& Ferrarese 2001).
Second, they use the empirical correlation between
the SMBH mass and the low-power emitting radio core for
E/S0 galaxies. They claim that the results estimated in both ways
are consistent. In Fig.5, the data estimated in the 
latter way is plotted.
As Fig.5 shows, the theoretical plot at $z=2.5$ already exceed the mass
estimate until now by more than one order of magnitude, thus, model B
is unreasonable.    
Since it is difficult to imagine that most of SMBHs evaporated, 
model B should be ruled out. 
The data is the current mass function of dormant SMBHs existing at
the center of E/S0 galaxies, but this is about 70\% of all SMBHs.
The amount of matter contained in the SMBHs probably increases, 
since there are still many QSOs at $z<2.5$, suggesting that accretion 
to the SMBHs continues. However, the number density of bright QSOs 
rapidly decreases towards $z \sim 0$, and the current number
density is about 1/100 of that of $z \sim 2.5$.     
 
\subsection{Formation epoch of SMBHs}

In this section, we consider the formation epoch of SMBHs.
As seen above, SMBH mass function can clearly discriminate model B 
from others, but
cannot break the degeneracy between A and C.
This is because the mass fraction, 
$\epsilon \equiv M_{\rm BH}/M_{\rm halo}$, 
is comparable in models A and C at redshifts
where we are concerned with, though in model A $\epsilon$
depends on the formation epoch of dark halos, $z_f$.
Under the CDM scenario, leading to hierarchical 
structure formation, 
the more massive halos forms at later times. 
Therefore, the formation epoch of one SMBH of fixed $M_{\rm BH}$ is different
among the models because of different $\epsilon$'s.         
 
Here, we calculate the probability that one SMBH of $M_{\rm BH}$ 
existing at $z=3.0$ had formed between $t_1({\rm Myr})$ and
$t_2 ({\rm Myr})$, where $t(z) > t_2 > t_1$. 
This quantity is described as following. 
\begin{equation}
p(t_1,t_2|M_{\rm BH},z) =
\frac{\int_{z_1}^{z_2}
d^2 N_{\rm form}/dzdM_{\rm BH} (M_{\rm BH},z_f,z) \ dz_f}
{d\Phi/dM_{\rm BH} (M_{\rm BH},z)}.
\end{equation}
Here, $z_1$ and $z_2$ is the redshift corresponding to time,
$t_1 ({\rm Myr})$ and $t_2 ({\rm Myr})$.
Furthermore, the same probability for BHs whose mass is more 
than $M_{\rm BH}$ is 
\begin{equation}
P(t_1,t_2| > M_{\rm BH},z) 
= \frac{
\int_{M_{\rm BH}}^{\infty} p(t_1,t_2|M_{\rm BH}',z) \ 
                d\Phi/dM_{\rm BH}(M_{\rm BH}',z) \ dM_{\rm BH}'}
{\int_{M_{\rm BH}}^{\infty}
               d\Phi/dM_{\rm BH} (M_{\rm BH}',z) \ dM_{\rm BH}'}.
\end{equation}
Fig.6 represents calculated $P(t_1,t_2| > 10^8 M_\odot,3.0)$.
This shows that in model C the fraction of SMBHs which were formed later 
is larger than in models A and B.
This difference is interpreted as follows.
As mentioned above, the QSO lifetime and the ratio of 
the SMBH mass to the halo mass are different among three models.
In models A,C, and B, in turn, $t_Q$ gets longer in order to 
fit the observed optical LFs at $2.2 < z < 4.4$. 
On the other hand, $M_{\rm BH}/M_{\rm halo}$ gets
smaller in the same turn(see Fig.4). 
But in model A, there is the redshift dependence on the ratio; 
$M_{\rm BH}/M_{\rm halo} \propto M_{\rm halo}^{2/3} (1+z_f)^{5/2}$.
That is, at higher redshifts, $M_{\rm BH}/M_{\rm halo}$ becomes larger
than model C. 
Since the less massive halos form earlier,                       
under the hierarchical structure formation scenario with the CDM   
cosmology, our results are consistent with this picture.         
In model A, as mentioned above, the growth of SMBHs is controled by 
the feedback from QSOs. This scenario derives the relation 
$M_{\rm BH} \propto M_{\rm halo}^{5/3} (1+z_f)^{5/2}$.
Especially, the dependence on $z_f$ is characteristic, and we need to
check this to ascertain the scenario. 
The formation epoch of SMBHs may be effective to confirm this $z_f$ depndence.

However, even if SMBHs exist at the centers of galaxies in an 
inactive state, as in the nearby universe, it is difficult to 
estimate the SMBH mass at the present day. 
To resolve these problems, we need to
investigate the host galaxies of inactive SMBHs at these redshifts.
Ridgway et al.(2001) note that the properties of 
the QSO host galaxies at $z \sim 2-3$, such as size and magnitude, are the 
same as those of the Lyman-break galaxies (e.g. Steidel et al. 1996) 
which existed at $z \sim 3$. 
On the other hand, Shapley et al.(2001) investigates about 100
Lyman-break galaxies and apply the population synthesis model for each
galaxy. They estimate that the elapsed time since the star formation 
event, $t_{\rm sf}$, covers from several ten Myr to 1 Gyr.
Especially, no less than 40\% of samples are $t_{\rm sf} > 500 {\rm
Myr}$.  
In our QSO scenario, every halo experiences the QSO phase
during $t_Q$, then these Lyman-break galaxies
can be the dead QSOs if $t_Q < 100 {\rm Myr}$.
Since many authors point the connection between the 
starburst and AGN (e.g. Aretxaga et al.1998), we may use $t_{\rm
sf}$ as the lifetime of SMBHs.

\section{Summary and Discussion}

In this paper, we consider the simple model for QSO LFs at high redshifts,
$z \sim 4.4,3.0$, and 2.6. 
We can reproduce the optical LFs at these redshifts
but with different models.
We thus considered how to constrain the models;
and how to determine observationally
QSO lifetime, $t_Q$, and the relation between the BH mass and the 
halo mass at high redshifts. 

We calculated the mass function of SMBHs at $z=2.5$.
The mass of the host halos, $M_{\rm halo}$, for a fixed mass
$M_{\rm BH}$ is different in each model, so the mass function of SMBHs
is different. The mass fraction $\epsilon \equiv M_{\rm BH}/M_{\rm halo}$
is smaller in model than in models A and C, and the less massive halos
has the larger population.
Therefore, the number density of SMBHs of the same mass is larger in
model B than in models A and C.
Compared with the current mass function, model B, in which $t_Q$ is 
comparatively shorter ($\sim 10^6$yr), should be ruled out,
since the calculated mass function based on model B at $z=2.5$ 
already exceeds the current value by more than one order of
magnitude. 
Furthermore, considering that there are still a lot of QSOs at $z \leq 2.5$,
we conclude that at least $t_Q \geq 10^7 {\rm yr}$ is needed.
Here, we note that this constraint can become even stronger with other
model assumptions.
For example, we assumed that each halo necessarily possesses
one SMBHs. If not all halos had SMBHs, the number density of QSOs decreases
with the same parameters. To fit observed LFs, therefore, we need 
larger $M_{\rm BH}/M_{\rm halo}$ or larger $t_Q$. In both cases,
the lower limit of $t_Q$ derived from the BH mass function becomes
large. It is the same case when most of QSOs shine at sub-Eddington 
luminosity. 

Next, we calculated the distribution of the formation epoch of SMBHs
existing at $z=3.0$. Formation epoch is also different in each model
because of different $\epsilon$. The key fact is that
less massive halos form at higher redshifts under the 
hierarchical structure formation in the CDM cosmology.
In model C, the mass of the halos corresponding to a fixed
BH mass is smallest at high redshifts, and hence the 
fraction of SMBHs forming at lower redshifts is largest.
We simply discussed that the information atout the formation eopch
of SMBHs is possible to prove model A.

{\acknowledgements
T.H. thanks Prof. Shin Mineshige for careful reading and useful discussion.
T.H. are also grateful to Dr.T.Kawaguch and  Dr.K.Yoshikawa for some 
hints and suggestions. 
}

\clearpage
\begin{figure}
\plotone{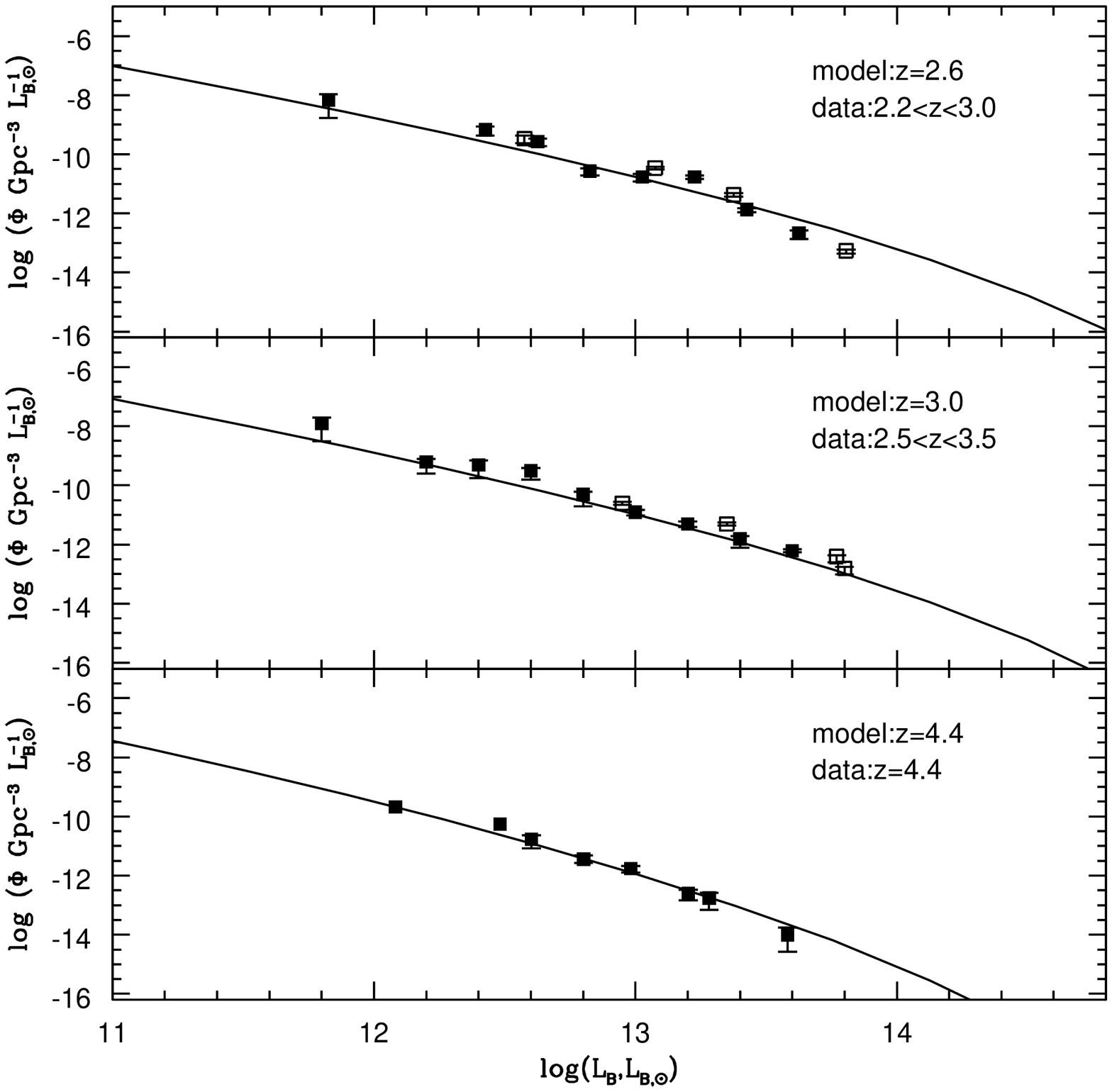}
\figcaption{Calculated QSO luminosity functions at high redshifts
based on the theoretical model A.
We use the data by Pei(1995) at $z=2.6$ and $z=3.0$, and
the data by Kennefick et al.(1996), Schmidt et al.(1995) and 
Kennefick et al.(1995) at $z=4.4$. The model parameters are
$t_Q = 6 \times 10^7$ yr, and $M_{\rm BH} = 1.58 \times 10^9 M_\odot 
(v_{\rm halo} / 500 {\rm km \ s}^{-1})^5$.}
\end{figure}

\clearpage
\begin{figure}
\plotone{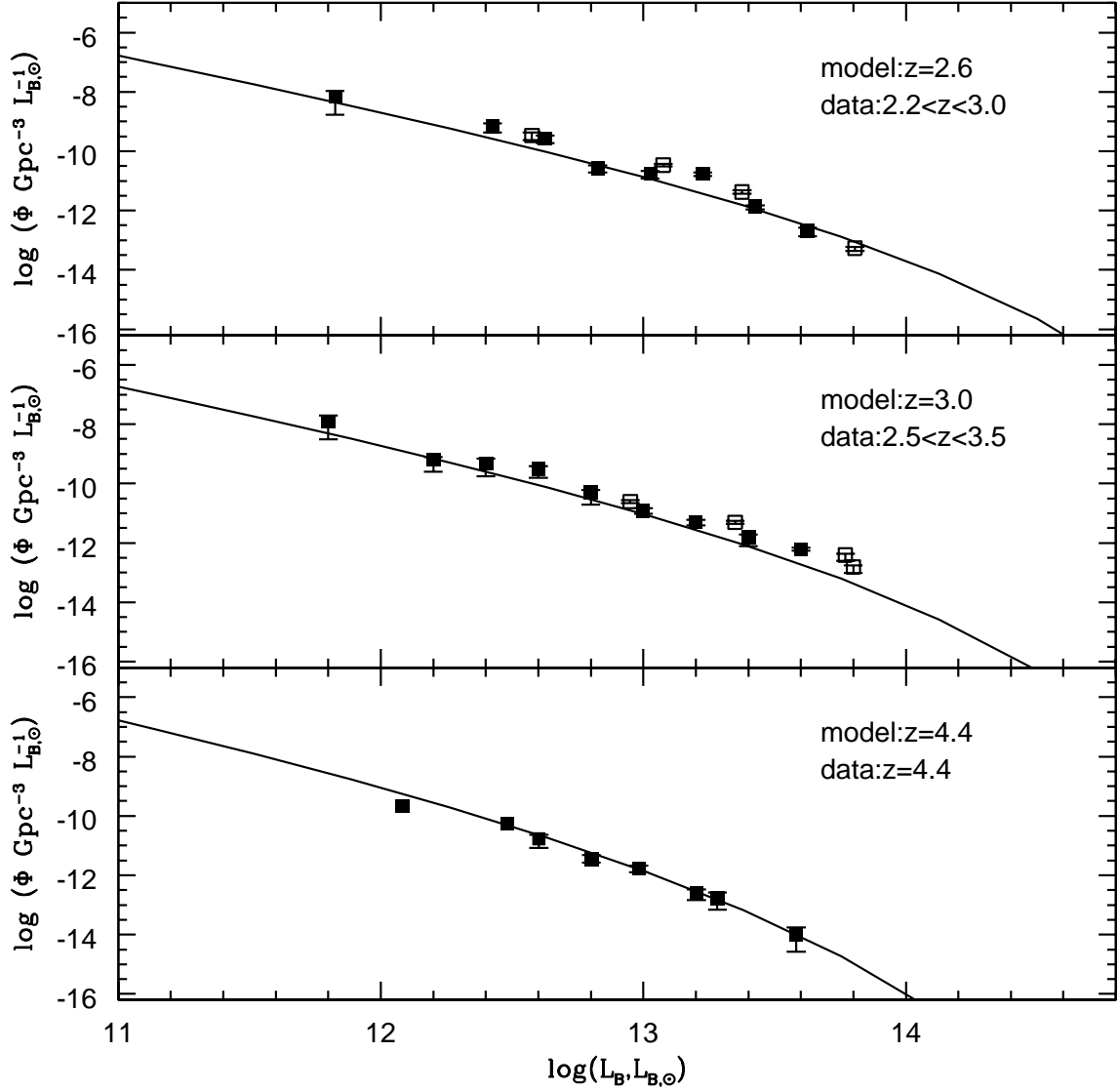}
\figcaption{Same as Fig.1 but for model B. 
The model parameters are
$t_Q = 10^6$ yr, and 
$\epsilon = M_{\rm BH}/M_{\rm halo}=1.26 \times 10^{-3}$.}
\end{figure}
\clearpage

\begin{figure}
\plotone{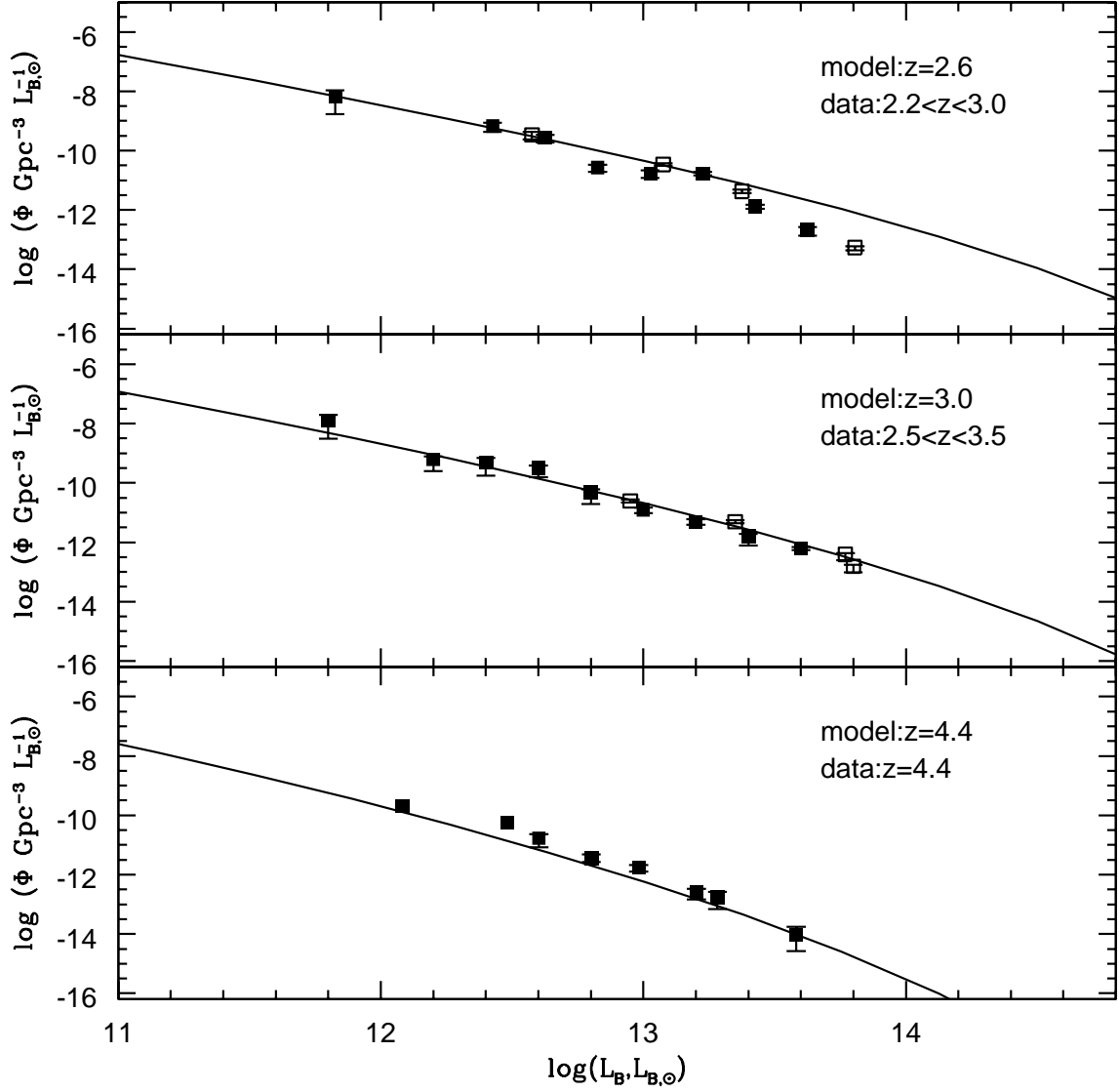}
\figcaption{Same as Fig.1 but for model C. The model parameters are
$t_Q = 10^7$ yr and $M_{\rm BH} = 7.13 \times 10^{-13} \times 
M_{\rm halo}^{5/3}$.}
\end{figure}
\clearpage

\begin{figure}
\plotone{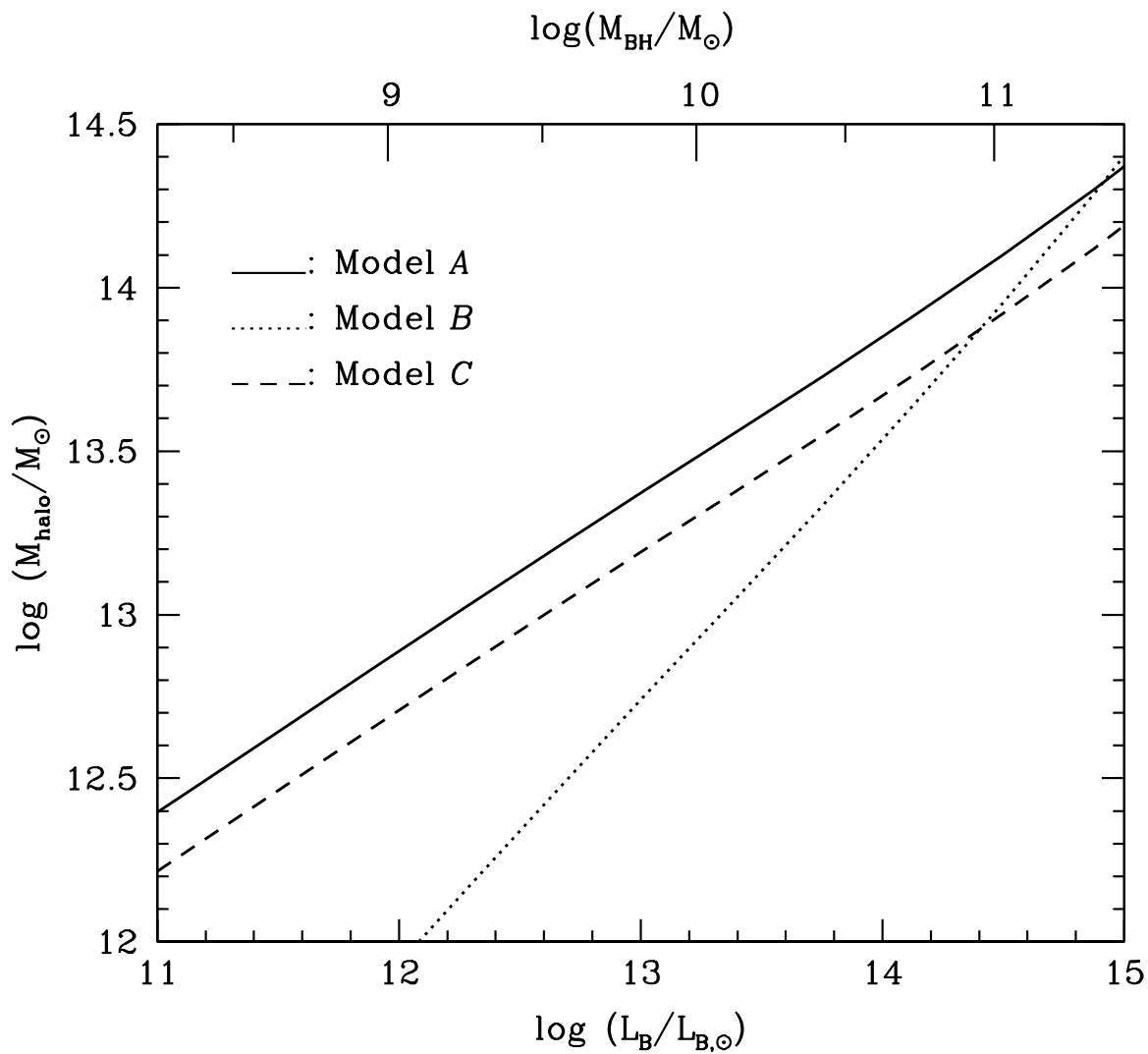}
\figcaption{Dominant black hole mass(top horizontal axis) and halo mass 
(bottom) calculated so as to reproduce QSO LFs at $z = 3$. 
The corresponding mass scale of BH is the same in these models,
while the mass scale of halos is different in 
model A(solid line), model B(dotted line), and model C(broken line).
This difference leads to the different $M_{\rm BH}/M_{\rm halo}$.}
\end{figure}
\clearpage

\begin{figure}
\plotone{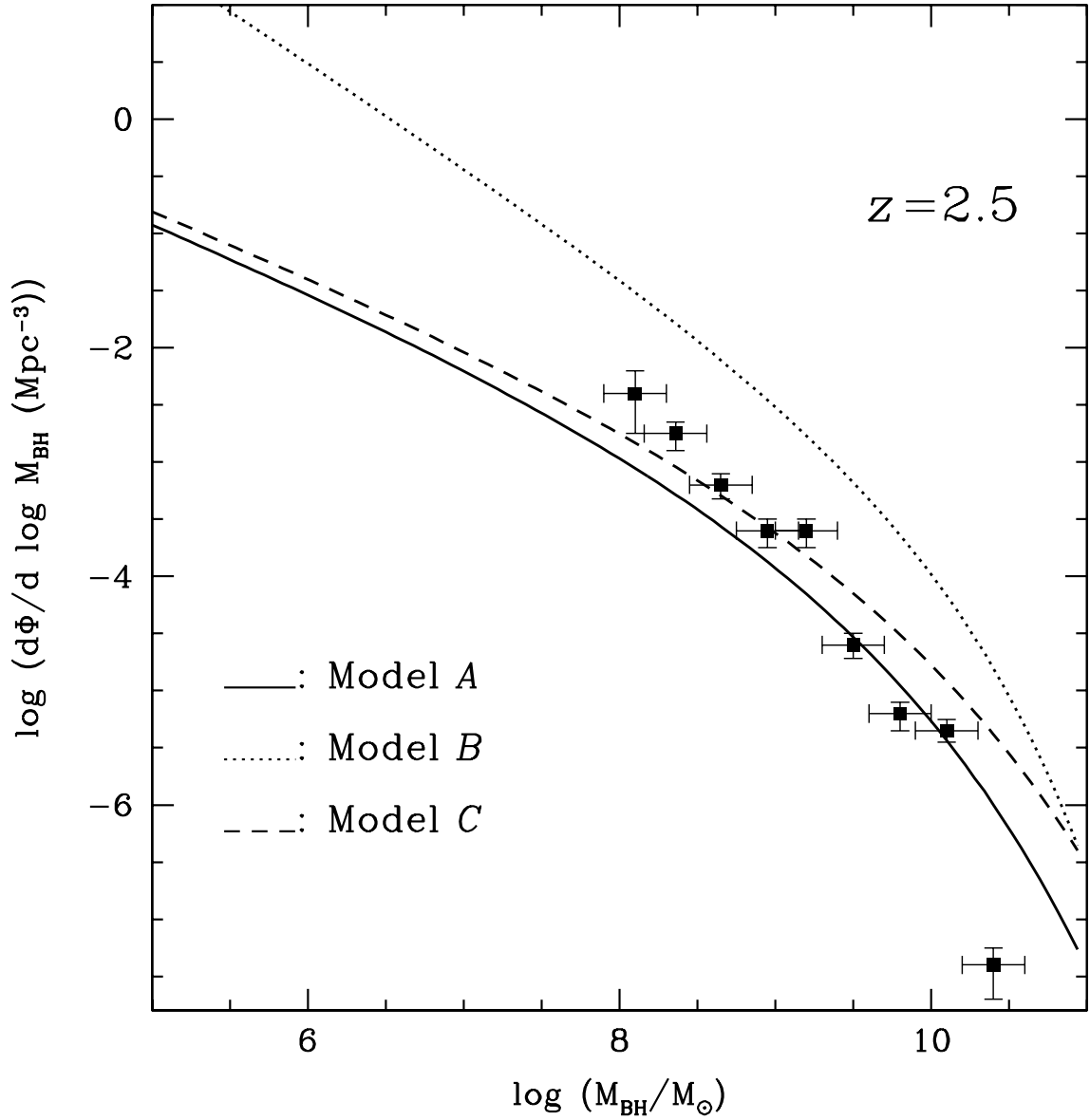}
\figcaption{Calculated black hole mass function at $z=2.5$.
The solid line represents model A,
the dotted line represents model B, and the broken line model C. 
The data point represents the current mass function by Salucci et al.(1999).}
\end{figure}

\clearpage
\begin{figure}
\plotone{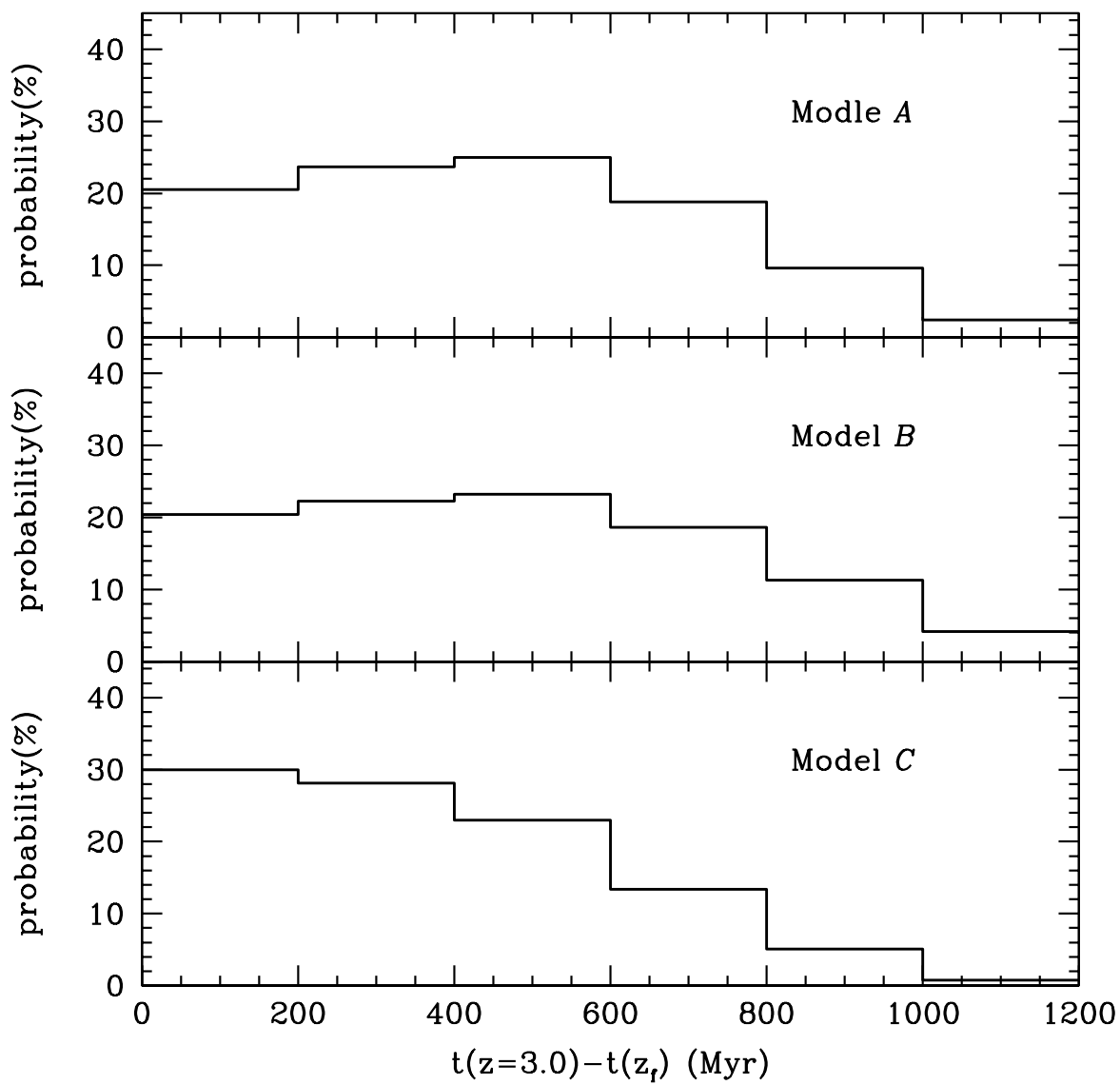}
\figcaption{The probability distribution of the formation epoch
of SMBHs of $M_{\rm BH} > 10^8 M_\odot$ 
existing at $z=3.0$, $P(t_1,t_2 | > 10^8 M_\odot, 3.0)$, 
based on model A(top), model B(middle),
and model C(bottom). Here, we set $t_2=t_1 + 200 ({\rm Myr})$. }
\end{figure}

\end{document}